\documentclass[a4paper,oneside,final,notitlepage,onecolumn,11pt]{article}
\usepackage{amsfonts}
\usepackage{epsf}
\usepackage{t1enc}
\usepackage[latin2]{inputenc}
\usepackage{graphicx}

\DeclareFontFamily{OT1}{rsfs}{} \DeclareFontShape{OT1}{rsfs}{m}{n}{
<-7> rsfs5 <7-10> rsfs7 <10-> rsfs10}{}
\DeclareMathAlphabet{\mycal}{OT1}{rsfs}{m}{n}

\DeclareFontFamily{OT1}{rsfs}{} \DeclareFontShape{OT1}{rsfs}{m}{n}{
<-7> rsfs5 <7-10> rsfs7 <10-> rsfs10}{}
\DeclareMathAlphabet{\mathscr}{OT1}{rsfs}{m}{n}

\def\eeepsilon{{\epsilon}{\hskip-.1399cm\epsilon}}

\begin{document}

\newtheorem{theorem}{Theorem}[section]
\newtheorem{lemma}{Lemma}[section]
\newtheorem{proposition}{Proposition}[section]
\newtheorem{corollary}{Corollary}[section]
\newtheorem{conjecture}{Conjecture}[section]
\newtheorem{example}{Example}[section]
\newtheorem{definition}{Definition}[section]
\newtheorem{remark}{Remark}[section]
\newtheorem{exercise}{Exercise}[section]
\newtheorem{axiom}{Axiom}[section]
\newtheorem{condition}{Condition}[section]
\renewcommand{\theequation}{\thesection.\arabic{equation}} 

\author{\small 
Istv\'{a}n R\'{a}cz
\thanks{
email: iracz@rmki.kfki.hu}
\\ 
\small  RMKI, H-1121 Budapest, Konkoly Thege Mikl\'os \'ut 29-33, Hungary 
}

\date{\small \today}

\title{{\bf On the topology of untrapped surfaces }}

\maketitle

\begin{abstract} 
Recently a simple proof of the generalizations of Hawking's black hole
topology theorem and its application to topological black holes for higher
dimensional ($n\geq 4$) spacetimes was given \cite{rnew}. By applying the
associated new line of argument it is proven here that strictly stable
untrapped surfaces possess exactly the same topological properties
as strictly stable {\it marginally outer trapped surfaces} (MOTSs) are known
to. In addition, a quasi-local notion of outwards and inwards
pointing  spacelike directions---applicable to untrapped and marginally
trapped  surfaces---is also introduced. 
\end{abstract}

\medskip 


\parskip 5pt

\section{Introduction}
\setcounter{equation}{0}

Hawking's black hole topology theorem \cite{hawk} plays a key role in
$4$-dimensional black hole physics from the beginning of the 70's. By making
use of a variant of Hawking's argument, almost three decades later, in the
late 90's, Gibbons \cite{Gib} and Woolgar \cite{wo} could also characterize
the so-called ``topological black hole'' spacetimes---to which Hawking's
original argument do not apply---by deriving a genus dependent lower bound for
the entropy of these black holes.  

Motivated by the considerable increase of interest in higher
dimensional black hole configurations of string theory and other
generalizations of Einstein theory of gravity, during the last couple of years
Galloway and his collaborators (see Refs.\,\cite{cg,ga1,ga2,ga3}) provided
important generalizations of Hawking's  \cite{hawk} black hole topology
theorem, and also that of Gibbons' \cite{Gib} and Woolgar's \cite{wo} results
to higher ($n\geq 4$) dimensional Einstein's theory of gravity where
considerations were restricted exclusively to 
marginal surfaces,
more precisely, to MOTSs. 

In our previous paper \cite{rnew} a simple self-contained proof of these
recent generalizations has been derived and it was also proved that these
results are valid in a more generic context, i.e., for a much higher variety
of theories, than anticipated before. Surprisingly, the applied new line of
argument {has} shown itself to be even more effective. More
specificly, according to 
the main result of the present paper,  the assertion of Theorem\,4.1 of
\cite{rnew} remains intact whereas the set of strictly stable MOTSs considered
therein is  subtended by the set of strictly stable untrapped surfaces. In
order to make the significance of this result to be more transparent, let us
mention that compact orientable MOTSs with no boundary are expected to
represent the boundary of black hole regions \cite{AMS} while untrapped
surfaces of the same type are thought to fill up the entire of the exterior
regions. Thereby the untrapped surfaces are, in fact, more common than
MOTSs. This {becomes} to be more manifest if one thinks
of the fact that they 
are also present in spacetimes which do not even have a black hole region. 


\section{Preliminaries}
\setcounter{equation}{0}

Since our argument does apply to any metric theory of gravity within this
paper, likewise in \cite{rnew}, a spacetime is supposed to be represented by a
pair $(M,g_{ab})$, where $M$ is an $n$-dimensional ($n\geq4$), smooth,
paracompact, connected, orientable manifold while  $g_{ab}$ is a smooth
Lorentzian metric of signature $(-,+,\dots,+)$ on $M$. We assume that
$(M,g_{ab})$ is time orientable and that a time orientation has been fixed. No
use {of} any sort of field equation concerning the spacetime metric
or the matter content will be {made}. Instead, only the
following generalized form of {\it dominant energy condition} will be
applied. As in \cite{rnew}, a spacetime $(M,g_{ab})$ is said to satisfy the
{\it generalized dominant energy condition} if there exists a smooth real
function $f$ on $M$ such that for all future directed timelike vector $t^a$
the combination $-[{G^a}_bt^b+f\,t^a]$ is a future directed timelike or null
vector, where $G_{ab}$ denotes the Einstein  tensor $R_{ab}-\frac{1}{2} g_{ab}
R$.  It is straightforward to justify (see, e.g., \cite{rnew} for details)
that in Einstein's theory of gravity the generalized dominant energy condition
holds, with the choice $f=\Lambda$, if and only if the energy-momentum tensor,
$T_{ab}$, satisfies the standard form of the dominant energy condition. 


As indicated above, {\it untrapped} surfaces of $n\geq 4$ dimensional
spacetimes will be at the center of our concern in this paper. In providing
their definition start with a smooth orientable $(n-2)$-dimensional compact
manifold $\mathscr{S}$ with no boundary in an $n$-dimensional spacetime
$(M,g_{ab})$. Let $\ell^a$ and $n^a$ be smooth future and past directed null
vector fields on  $\mathscr{S}$, respectively, which are scaled such that
$n^a\ell_a=1$, and that are also normal to $\mathscr{S}$, i.e.,
$g_{ab}\ell^aX^b|_{\mathscr{S}}=g_{ab}n^aX^b|_{\mathscr{S}}=0$ for any vector
field  $X^a$ tangent to $\mathscr{S}$. Note that these conditions
ensure 
that neither $\ell^a$ nor $n^a$ vanishes on $\mathscr{S}$. Consider then the
null hypersurfaces {$\mathcal{L}$ and $\mathcal{N}$} generated by
geodesics starting on $\mathscr{S}$ with 
tangent $\ell^a$ and $n^a$. By choosing synchronized affine parameterizations
to these null geodesic curves the vector fields  $\ell^a$ and $n^a$ extend
{to $\mathcal{L}$ and $\mathcal{N}$},
respectively. These hypersurfaces are
smooth in a 
neighborhood of $\mathscr{S}$, and also they are smoothly foliated by the
level surfaces---which are  $(n-2)$-dimensional compact manifolds homologous
to $\mathscr{S}$---of the applied synchronized affine parametrizations.
Denote by ${\eeepsilon}_q$ the volume element associated with the metric,
$q_{ab}$, induced on these $(n-2)$-dimensional surfaces. Then the null
expansions $\theta^{(\ell)}$ and $\theta^{(n)}$ with respect to $\ell^a$ and
$n^a$ are defined as
\begin{equation}\label{exp}
\pounds_\ell\,\eeepsilon_q=\theta^{(\ell)}\,{\eeepsilon}_q
{ \ \ {\rm and}\ \ }
\pounds_n\,{\eeepsilon}_q= \theta^{(n)}\,{\eeepsilon}_q\,,   
\end{equation}  
where $\pounds_\ell$ and $\pounds_n$ denote the Lie
derivative{s} with respect to the null vector fields $\ell^a$ and
$n^a$, respectively.

According to the original definition of Penrose \cite{P} a $2$-dimensional
surface $\mathscr{S}$, in a $4$-dimensional spacetime, is considered to be
future or past trapped if both of the future or past directed null geodesic
congruences orthogonal to $\mathscr{S}$ are converging at $\mathscr{S}$,
respectively. Thereby, an $(n-2)$-dimensional surface $\mathscr{S}$ is called
to be {\it trapped} if the null expansions $\theta^{(\ell)}$ and
$\theta^{(n)}$ are such that one of them is non-negative while the other is
non-positive throughout $\mathscr{S}$. To separate the limiting case it is
assumed that neither of these expansions vanishes identically on
$\mathscr{S}$. Accordingly, an $(n-2)$-dimensional surface $\mathscr{S}$ is
called to be {\it untrapped} if the null expansions $\theta^{(\ell)}$ and
$\theta^{(n)}$ are both non-negative or non-positive throughout $\mathscr{S}$
and neither of them vanishes identically there. In the limiting case, i.e.,
whenever either of the expansions gets to be identically zero $\mathscr{S}$ is
called to be {\it marginal}. 


It is worth keeping in mind that there is a much higher variety of
$(n-2)$-dimensional surfaces than the ones covered by the above three
categories. As an immediate example one may think of a surface yielded by a
generic deformation of a marginal surface $\mathscr{S}$, one of the null
expansions of which, say $\theta^{(\ell)}$, is identically zero on
$\mathscr{S}$ originally, but which may be altered such that $\theta^{(\ell)}$
changes sign after the deformation is performed. 

What {distinguishes} marginal and untrapped
surfaces is that a 
meaningful quasi-local concept of outwards and inwards directions can be
associated with these surfaces which---as it immediately follows from the
details of the construction below---cannot be done in case of generic
$(n-2)$-dimensional surfaces. In introducing these notions let us consider a
marginal or an untrapped surface $\mathscr{S}$ and assume that
$\theta^{(\ell)}\geq 0$ and $\theta^{(n)}\geq 0$. [Note that these
inequalities may be assumed to be satisfied, without loss of generality, since
if they did not hold we could apply the transformation $\ell^a\rightarrow
\ell'^a=-n^a$ and $n^a\rightarrow n'^a=-\ell^a$ to ensure them.]  Consider now
a vector field $Z^a$ that is given by {a} linear
combination {of the form} $Z^a=A \ell^a +B
n^a$,  where the coefficients, $A$ and $B$, are smooth
functions 
on $\mathscr{S}$. It is straightforward to see that for any particular choice
of $A$ and $B$ the vector field $Z^a$ is smooth and spacelike everywhere on
$\mathscr{S}$ whenever $A$ and $B$ are both positive or negative throughout
$\mathscr{S}$. 

In order to justify that a meaningful quasi-local concept of outwards and
inwards directions may be adequately associated with these type of nowhere
vanishing spacelike vector fields fix the coefficients $A$ and $B$ and
consider the spacelike geodesics staring on $\mathscr{S}$ with tangent
$Z^a$. Denote by $z$ and $\tilde Z^a$ a synchronized affine parameter and the
associated tangent field along these geodesics, respectively. Consider then
the one-parameter family of surfaces $\mathscr{S}_z$ yielded by the
Lie-transport of $\mathscr{S}$ with respect to $\tilde Z^a$. It is said then
that the vector field $Z^a$, defined on $\mathscr{S}$, points {\it outwards}
or {\it inwards} if the total variation $\delta_Z\mathcal{A}=\frac{{\rm d}
  \mathcal{A}(\mathscr{S}_z)}{{\rm  d}z}\vert_{z=0}$ of the ``area''
$\mathcal{A}(\mathscr{S})=\int_{\mathscr{S}}\,\eeepsilon_q$, in the direction
of $Z^a$, is positive or negative, while $\delta_Z\mathcal{A}(\mathscr{S}')\geq
0$ or $\delta_Z\mathcal{A}(\mathscr{S}')\leq 0$ for any subset
$\mathscr{S}'\subset\mathscr{S}$, respectively. To see that the above selected
spacelike vector fields do always have a definite orientation recall that for
any subset 
$\mathscr{S}'\subset\mathscr{S}$
\begin{equation}\label{av00}
\delta_Z\mathcal{A}(\mathscr{S}')= \int_\mathcal{\mathscr{S}'}
    \pounds_{\tilde Z}\,\eeepsilon_q=
    \int_\mathcal{\mathscr{S}'} \left[ A\,\theta^{(\ell)}+B\,\theta^{(n)}\right]
    \,\eeepsilon_q\, 
\end{equation} 
which has the appropriate sign whenever the coefficients $A$ and $B$  are both
positive or negative  throughout $\mathscr{S}$, respectively. If both
$\theta^{(\ell)}$ and $\theta^{(n)}$ vanish identically on $\mathscr{S}$ the
above quasi-local concept of outwards and inwards directions become degenerate
in accordance with the fact that then $\mathscr{S}$ is a minimal surface.


In deriving our main results, likewise in \cite{rnew}, we shall also apply a
stability assumption. In formulating it recall first that the null vector
fields $\ell^a$ and $n^a$ on $\mathscr{S}$ are not uniquely determined by the
conditions we have imposed. In fact, together with $\ell^a$ and $n^a$ the null
vector fields 
\begin{equation}\label{boost}
\ell'^a=e^{-v}\,\ell^a\ \ \ {\rm and}\ \ \ \ n'^a=e^v\,n^a\,
\end{equation}  
are also suitable, where $v: \mathscr{S} \rightarrow \mathbb{R}$ is an
arbitrary smooth
function on $\mathscr{S}$. It is worth emphasizing,\,however,\,that the
signs of $\theta^{(\ell)}$ and $\theta^{(n)}$,\,and,\,in turn,\,the notion of
trapped, untrapped and marginal surfaces,\,along with the above defined
quasi-local notion of outwards and inwards directions, are intact under such a
positive rescaling.

An untrapped surface $\mathscr{S}$ is called 
to be {\it  strictly stable} if there exists a boost transformation of the
form (\ref{boost}) such that $(\pounds_{n'}\,\theta^{(\ell')}+\theta^{(\ell')}
\,\theta^{(n')}) \vert_{\mathscr{S}}\geq 0$ for the vector fields $\ell'^a$
and $n'^a$,\,and such that the inequality is strict somewhere on
$\mathscr{S}$.\,In the discussion part we shall return to the interpretation
of this---apparently technical but in its fundamental nature purely
geometrical---requirement.\,{It is worth mentioning though that,\,e.g.,\,all the metric spheres of
  the Minkowski spacetime or that of a Schwarzschild spacetime with radius
  $r>2M$ are strictly stable untrapped surfaces according to the above
  definition.}

As it was emphasized {earlier} our aim in this paper is
to provide a topological 
characterization of {strictly stable} untrapped surfaces. It is
well-known that in this type of 
characterization of surface{s} of
dimension $s=n-2\geq 3$, in an 
$n$-dimensional spacetime, the sign of the Yamabe invariant plays a
distinguished role. For instance, whenever it is possible to demonstrate that
the Yamabe invariant of $\mathscr{S}$ is positive, as it follows
from the 
remarkable results of Gromov and Lawson \cite{GL}, $\mathscr{S}$ cannot
carry a metric of non-positive sectional curvature which, in turn, raises
restriction{s} on the topology of $\mathscr{S}$. 

In recalling the notion of the Yamabe invariant consider first the conformal
class $[q]$  of Riemannian metrics on $\mathscr{S}$ determined by $q_{ab}$,
i.e., $[q]$ consist of those Riemannian metrics $\hat q_{ab}$ on $\mathscr{S}$
which can be given as positive function times $q_{ab}$. Then, the conformal
invariant Yamabe
constant $Y(\mathscr{S},[q])$, associated with the conformal class $[q]$, is
given as 
\begin{equation}\label{y1}
Y(\mathscr{S},[q])= \inf_{\hat q\in[q]}\frac{\int_{\mathscr{S}}R_{\hat
    q}\,\eeepsilon_{\hat q}} 
{\left(\int_{\mathscr{S}}\eeepsilon_{\hat q}\right)^{\,\,\frac{s-2}{s}}}\,,
\end{equation} 
where $R_{\hat  q}$ denotes the scalar curvature associated with the
Riemannian metric $\hat q_{ab}$. The Yamabe invariant
$\mathcal{Y}(\mathscr{S})$ is defined then as the supremum of the Yamabe
constants associated with $\mathscr{S}$, i.e., $\mathcal{Y}(\mathscr{S}) =
\sup_{[q]}Y(\mathscr{S},[q])$. 

It is worth recalling that according to important results of Aubin and Schoen
the Yamabe invariant $\mathcal{Y}(\mathscr{S})$ is known to be bounded
from above by the Yamabe constant of a sphere of dimension $s=n-2\geq 3$ with
its standard metric. It also immediately follows from (\ref{y1}) that in case
of a  $2$-surface the Yamabe constant reduces to {$4\pi$ times}
the Euler characteristic 
$\chi_{_{\mathscr{S}}}$ of $\mathscr{S}$, i.e., $Y(\mathscr{S},[q])=4\pi
\chi_{_{\mathscr{S}}}$,  for any conformal class $[q]$ of Riemannian metrics
on $\mathscr{S}$. Thereby, in virtue of its definition, the Yamabe invariant
itself is also equal to $4\pi$ times the Euler characteristic of
$\mathscr{S}$. 


\section{The main result}
\setcounter{equation}{0}

Now, by making use of the above recalled notions, our main result is
formulated as. 

\begin{theorem}\label{HBT}  Let $(M,g_{ab})$ be a spacetime of dimension
$n\geq 4$ in a metric theory of gravity. Assume that the generalized dominant
energy condition, with smooth real function $f:M\rightarrow \mathbb{R}$, holds
and that $\mathscr{S}$ is a strictly stable untrapped surface in $(M,g_{ab})$.
\begin{itemize}
\item[(1)] If $f\geq 0$ on $\mathscr{S}$ then $\mathscr{S}$ is of positive
  Yamabe type, 
  i.e., $\mathcal{Y}(\mathscr{S})>0$. 
\item[(2)] If $\mathcal{Y}(\mathscr{S})<0$ and $f^{_{\mathscr{S}}}_{min}<0$,
  where $f^{_{\mathscr{S}}}_{min}$ denotes the minimal value of $f$ on
  $\mathscr{S}$,  then  
\begin{equation}\label{gib2}
\mathcal{A}(\mathscr{S})\geq
\left(\frac{|\mathcal{Y}(\mathscr{S})|}{2|f^{_{\mathscr{S}}}_{min}|}\right)
^{\,\,\frac s2}\,.  
\end{equation}
\end{itemize}     
\end{theorem}   \textbf{Proof:}  In justifying the above assertions {the
  following 
generalization} of the argument of \cite{rnew} will be applied. In order
to recall some of the basics of the associated simple geometric setup start
with the smooth null hypersurface $\mathcal{N}$ spanned by 
the $(n-2)$-parameter congruence of null geodesics starting at $\mathscr{S}$
with tangent $n^a$. Denote by $u$ the affine parameter along these geodesics
that is synchronized such that $u=0$ on  $\mathscr{S}$ and by $n^a$ the
tangent field $(\partial/\partial u)^a$ on $\mathcal{N}$. The $u=const$
cross-sections, $\mathscr{S}_u$, provides then a smooth foliation of
$\mathcal{N}$. Denote by $\ell^a$ the unique future directed null vector field
on $\mathcal{N}$ defined by requiring that $g_{ab}n^a\ell^b=1$, and that
$\ell^a$ is orthogonal to each $\mathscr{S}_u$. Choose $r$ to be the affine
parameter of the null geodesics determined by $\ell^a$ which are synchronized
such that $r=0$ on $\mathcal{N}$.  


Since $\ell^a$ is, by construction, smooth on $\mathcal{N}$ the null geodesics
starting with tangent $\ell^a$ on $\mathcal{N}$ do not meet  within a
sufficiently small open  ``elementary spacetime neighborhood'' $\mathcal{O}$
of $\mathscr{S}$. The function $u$ is extended then from $\mathcal{N}$ onto
$\mathcal{O}$ by requiring its value to be constant along the geodesics with
tangent  $\ell^a$. Then the vector fields $n^a$ and $\ell^a$, defined so far
only on $\mathcal{N}$, do also extend onto $\mathcal{O}$ such that the
relations $n^a=(\partial/\partial u)^a$ and $\ell^a=(\partial/\partial r)^a$
hold there which immediately guarantee that $n^a$ and $\ell^a$ commute on
$\mathcal{O}$. The elementary spacetime neighborhood $\mathcal{O}$ is smoothly
foliated then by the $2$-parameter family of $(n-2)$-dimensional  $u=const$,
$r=const$ level surfaces $\mathscr{S}_{u,r}$, furthermore,  the spacetime
metric in $\mathcal{O}$ takes the form
\begin{equation}\label{metric}
g_{ab}=2\,\left(\nabla_{(a}r - r\,\alpha\,\nabla_{(a}u - r\,\beta_{(a}\right)
\nabla_{b)}u +\gamma_{ab}\,, 
\end{equation} where $\alpha$, $\beta_a$ and $\gamma_{ab}$ are smooth 
fields on $\mathcal{O}$ such that $\beta_a$ and $\gamma_{ab}$ are orthogonal to
$n^a$ and $\ell^a$ \cite{hiw}. 

Then, by making use of this simple geometrical setup---in particular,
equations (3.3), (3.5), (3.6) and (3.7) of \cite{rnew} such that all the
terms proportional to $\theta^{(\ell)}$ are retained---the relation 
\begin{eqnarray}\label{GDEC2}
&&\hskip-.5cm\pounds_n\theta^{(\ell)}\vert_{\mathscr{S}}=
G_{ab}n^a\ell^b-\alpha\,\theta^{(\ell)}-\theta^{(\ell)}
  \theta^{(n)} 
+\frac12\left[R_q+D^a\beta_a-\frac12\beta^a\beta_a
  \right]\,
\end{eqnarray}
can be deduced. 

In proceeding note first that, according to the following lemma, the
second term on the right hand side of the previous equation drops out.
\begin{lemma}
The metric function $\alpha$ vanishes on $\mathcal{N}$.
\end{lemma}
\noindent\textbf{Proof:}{\ } In virtue of (\ref{metric}) we have that 
$n^an_a=-2\,r\,\alpha\,,$
which, in particular, implies that 
 the relation 
\begin{equation}
\ell^e\nabla_e\left(n^an_a\right)=\partial_r(-2\,r\,\alpha)=-2\,\alpha
\end{equation}
is satisfied on $\mathcal{N}$, i.e., whenever $r=0$. The assertion of our
lemma follows then from the fact that  
\begin{eqnarray}
&&\ell^e\nabla_e\left(n^an_a\right)=2\, n_a\, \ell^e\nabla_e n^a= 2\, n_a\,
n^e\nabla_e \ell^a 
= - 2\, \ell_a\, n^e\nabla_e n^a=0\,,
\end{eqnarray}
holds on $\mathcal{N}$, where $[n,\ell]^a=0$ and 
$n_a\ell^a=1$, along with the fact that $u$ was chosen to be an affine
parameter along the generators of $\mathcal{N}$, have been applied.  
\hfill\fbox{} 

In addition, since $-n^a$ and $\ell^a$ are both future directed null vector
fields on $\mathscr{S}$, and also the generalized dominant energy condition
holds the
inequality $G_{ab}n^a\ell^b+f\leq 0$ {is satisfied} on
$\mathscr{S}$. Finally, recall 
that $\mathscr{S}$ was assumed to be a strictly stable untrapped surface which
ensures that the null normals $n^a$ and $\ell^a$ may be assumed, without loss
of generality, to be such that
$\left(\pounds_n\theta^{(\ell)}+\theta^{(\ell)} \theta^{(n)}\right)
\vert_{\mathscr{S}}\geq 
0$, and also that $\pounds_n\theta^{(\ell)}+\theta^{(\ell)}
\theta^{(n)}> 0$ somewhere on $\mathscr{S}$.

Consequently, whenever $\mathscr{S}$ is a strictly stable untrapped surface
and the  generalized dominant energy condition is also satisfied then, in
virtue of (\ref{GDEC2}), the inequality    
\begin{equation}\label{GDEC3}
R_q+ D^a\beta_a - \frac12 \beta^a\beta_a \geq 2\,f\,
\end{equation} 
hold{s}, so that it is strict somewhere on 
$\mathscr{S}$. Since (\ref{GDEC3}) possesses exactly the same form as (3.12)
in \cite{rnew} from this point the assertions of Theorem\,\ref{HBT} may be
justified simply by repeating the corresponding part of the argument of
Section 3 of \cite{rnew}.  \hfill\fbox{} 



\section{Discussion}
\setcounter{equation}{0}

Let us return now to the interpretation of the stability condition we have
applied. To this end note first that the second variation
$\delta_n\delta_\ell\mathcal{A}=\frac{{\rm \partial^2}
  \mathcal{A}(\mathscr{S}_{u,r})}{ \partial u\,\partial r}\vert_{u=0,r=0}$ of
the area in the principal null
directions $\ell^a$ and $n^a$ reads as  
\begin{equation}\label{2nd}
\delta_n\delta_\ell\mathcal{A}=\int_\mathcal{\mathscr{S}} 
    \pounds_{n}\pounds_{\ell}\,\eeepsilon_q=
    \int_\mathcal{\mathscr{S}} \left[ \pounds_{n}\theta^{(\ell)}
      +\theta^{(\ell)}\theta^{(n)}\right] 
    \,\eeepsilon_q\, . 
\end{equation}   Note that
$\delta_n\delta_\ell\mathcal{A}=\delta_{-n}\delta_{-\ell}\mathcal{A}$ and that
$\delta_n\delta_\ell\mathcal{A}=\delta_\ell\delta_n\mathcal{A}$, where the
first equality is trivial while the second one follows from fact that $\ell^a$
and $n^a$ commute in $\mathcal{O}$. According to the above observations our
strict stability condition is equivalent to the existence of
principal null vector  fields $\ell^a$ and $n^a$ on $\mathscr{S}$ such that
the second variation $\delta_n\delta_\ell\mathcal{A}$ is positive while
$\delta_n\delta_\ell\mathcal{A}(\mathscr{S}')\geq 0$ for any
portion $\mathscr{S}'\subset \mathscr{S}$. Note that the strict stability
condition used in \cite{AMS,AMSn,cg,ga1,ga2,ga3,rnew}{, only in the
  context of MOTSs,} can also be seen to be
equivalent to these requirements.

\smallskip

To justify our last claim and also to provide another characterization of the
applied strict stability condition  recall that under the rescaling
(\ref{boost}) of the vector fields $\ell^a$ and $n^a$ on $\mathscr{S}$ the
metric function $\alpha$ and the form field $\beta_a$ transform as
$\alpha\rightarrow \alpha'=\alpha\,e^v$ and $\beta_a\rightarrow \beta'_a=
\beta_a + D_a v$.  Introducing then the notation $\psi=e^{-2v}$ and
$s_a=\frac12\beta_a$, it can be verified  (see also \cite{rnew} for more
details) that  (\ref{GDEC2}) takes the form 
\begin{eqnarray}\label{GDEC2a}
&&\hskip-.5cm\big(\big[\pounds_{
    n'}\theta^{(\ell')}+\theta^{(\ell')}\,\theta^{( n')}\big]\,\psi\big)
  \vert_{\mathscr{S}}= 
-D^aD_a\psi 
+2s^aD_a\psi\\ &&\hskip6.12cm+\left[\frac12\, R_q+G_{ab}n^a\ell^b
  + D^as_a-s^as_a \right]\psi\,,\nonumber
\end{eqnarray}  
where the vanishing of $\alpha$ on $\mathscr{S}$ has been
applied. By choosing then ``the variation vector field $v^a$'' to be $n^a$ the
right hand side of (\ref{GDEC2a}) coincides with the action of the ``stability
operator''  $L_v$, defined by equation (5) of \cite{AMSn}, on $\psi$.  Thereby
the right hand side of (\ref{GDEC2a}) determines a linear elliptic operator
of the form given by equation (10) of   \cite{AMSn}, and whence the arguments
of Sections 4 and 5 of  \cite{AMSn} can also be applied to the present
case. Accordingly, $\mathscr{S}$ is a strictly stable untrapped  surface iff
there exists a function $\psi\geq 0$, $\psi\not\equiv 0$ on $\mathscr{S}$ such
that $L_v\psi\geq 0$, $L_v\psi\not\equiv 0$ on $\mathscr{S}$, or equivalently
iff the principal eigenvalue of $L_v$ is positive.


{As the domain of outer communication (DOC) of a black hole spacetime is
  filled up with untrapped surfaces our main result has to have some
  connection with the topological censorship theorems which provide a
  topological characterization of DOCs of asymptotically flat or
  asymptotically locally anti-de Sitter spacetimes (see, e.g.,
  \cite{fsw,gsww}). To indicate the most important differences recall that the
  topological censorship theorems---that are also  known (see, e.g.,
  \cite{gsww}) to be valid in arbitrary dimension $n\geq 3$---are inherently
  global. They assert that the topology of a DOC is {\it simple} if the
  topology of the asymptotic region is  simple. As opposed to this our result
  is fully quasi-local. Thereby, it applies to untrapped surfaces regardless
  whether the underlying spacetime possesses an asymptotic region or
  not. Nevertheless,  it would be useful, not only in exploring some of the
  deeper connections, to find out whether a foliation of DOCs by strictly
  stable untrapped surfaces is always possible.} 


Finally, we would like to emphasize again that in virtue of Theorem\,\ref{HBT}
and Theorem\,4.1 of \cite{rnew} strictly stable MOTSs and untrapped surfaces do
possess the same topological properties whenever both type of surfaces exist
in a spacetime. In the particular case of a $4$-dimensional spacetime in
Einstein's theory of gravity with zero cosmological constant and with matter
satisfying the dominant energy condition strictly stable untrapped surfaces
must possess the topology of $2$-spheres regardless whether the underlying
spacetime contains a black hole region or no{t}. Nevertheless, it is
also  
important to keep in mind that since no use of field equations {has
  been} made 
anywhere in our analysis and the spacetime dimension
{has} also {been} kept
arbitrary the main results of this paper applies to any metric theory of
gravity, with dimension $n\geq 4$.

\section*{Acknowledgments}
This research was supported in part by OTKA grant K67942.


\begin{thebibliography}{99}

\bibitem{AMS} L. Andersson, M. Mars, and W. Simon: 
{\it Local existence of dynamical and trapping horizons},
 Phys. Rev. Lett. {\bf 95}, 111102 (2005) 

\bibitem{AMSn} L. Andersson, M. Mars, and W. Simon:
{\it Stability of marginally outer trapped surfaces and existence of
  marginally outer trapped tubes}, 
arXiv:0704.2889

\bibitem{cg} M. Cai and G. J. Galloway: 
{\it On the topology and area of higher-dimensional black holes}, 
Class. Quant. Grav. {\bf 18}, 2707-2718
  (2001) 

\bibitem{fsw}
  J.L.\,Friedman,\,K.\,Schleich\,and\,D.M.\,Witt: {\it Topological
    Censorship}, Phys.\,Rev.\,Lett.\,{\bf 71}, 
1486-1489 (1993); Erratum-ibid.\,{\bf 75} 1872 (1995)

\bibitem{ga1} G.J. Galloway and R. Schoen: 
{\it A generalization of
  Hawking's black hole topology theorem to higher dimensions},
  Commun. Math. Phys. {\bf 266}, 571-576 (2006) 

\bibitem{ga2} G.J. Galloway: 
{\it Rigidity of outer horizons and the topology of black holes}, 
{Commun. \,Anal. \,Geom. \,{\bf
    16},\,217-229\,(2008)}

\bibitem{ga3} G.J. Galloway and N. O'Murchadha:  
{\it Some remarks on the size of bodies and black holes}, 
Class. Quant. Grav. {\bf 25}, 105009 (2008)

\bibitem{gsww}
  G.J.\,Galloway,\,K.\,Schleich,\,D.M.\,Witt\,and\,E.\,Woolgar: {\it  The
    AdS\,/\,CFT correspondence conjecture and topological censorship},   
  Phys.\,Lett.\,B\,{\bf
  505},\,255-262\,(2001)   

\bibitem{Gib} G.W.\,Gibbons:\,Class.\,Quant.\,Grav.\,{\bf
  16},\,1677-1687\,(1999)  
{\it Some Comments on Gravitational Entropy and
    the Inverse Mean Curvature Flow}, 

\bibitem{GL} M. Gromov and H.B. Lawson: 
{\it Positive Scalar Curvature and the
  Dirac Operator on Complete Riemannian Manifolds}, 
Publ. Math. IHES {\bf 58}, 83-196 (1983); 

\bibitem{hawk}
  S.W.\,Hawking: \textit{Black holes in general relativity},
Commun.\,Math.\,Phys.\,\textbf{25},\,152-166 \,(1972) 

\bibitem{hiw}
  S.\,Hollands,\,A.\,Ishibashi\,and\,R.M.\,Wald:
{A higher
 dimensional stationary rotating black hole must be axisymmetric},
\,Commun.\,Math. Phys.\,{\bf
    271},\,699-722\,(2007) 


\bibitem{P} R. Penrose: 
{\it Gravitational collapse and space-time singularities}, 
Phys. Rev. Lett. {\bf 14} 54-59 (1965)

\bibitem{rnew} I. R\'acz: {\it A simple proof of the recent generalizations 
of Hawking's black hole topology theorem}, 
{Class.\,Quant.\,Grav.\,{\bf
    25},\,162001\,(2008)} 


\bibitem{wo} E. Woolgar: 
{\it Bounded area theorems for higher-genus black holes},   
Class. Quant. Grav. {\bf 16}, 3005-3012  (1999)


\end{thebibliography}
\end{document}